\documentclass[pra,twocolumn,superscriptaddress,showpacs]{revtex4-1}
\usepackage[english]{babel}
\usepackage[dvips]{graphicx}
\usepackage{amssymb}
\usepackage{amsmath}
\usepackage{color}
\begin{document}

\title{Emission spectrum of the driven nonlinear oscillator}

\author{Stephan Andr\'e}
\affiliation{Institut f\"ur Theoretische Festk\"orperphysik,
      Karlsruhe Institute of Technology, 76128 Karlsruhe, Germany}
\affiliation{DFG Center for Functional Nanostructures (CFN),
      Karlsruhe Institute of Technology, 76128 Karlsruhe, Germany}
\author{Lingzhen Guo}
\affiliation{Institut f\"ur Theoretische Festk\"orperphysik,
      Karlsruhe Institute of Technology, 76128 Karlsruhe, Germany}
\affiliation{Department of Physics, Beijing Normal University, Beijing 100875, China}
\author{Vittorio Peano}
\affiliation{Department of Physics and Astronomy, Michigan State University, East Lansing, Michigan 48824, USA}
\author{Michael Marthaler}
\affiliation{Institut f\"ur Theoretische Festk\"orperphysik,
     Karlsruhe Institute of Technology, 76128 Karlsruhe, Germany}
\affiliation{DFG Center for Functional Nanostructures (CFN),
     Karlsruhe Institute of Technology, 76128 Karlsruhe, Germany}
\author{Gerd Sch\"on}
\affiliation{Institut f\"ur Theoretische Festk\"orperphysik,
      Karlsruhe Institute of Technology, 76128 Karlsruhe, Germany}
\affiliation{DFG Center for Functional Nanostructures (CFN),
      Karlsruhe Institute of Technology, 76128 Karlsruhe, Germany}
\date{\today}

\begin{abstract}
Motivated by recent ``circuit QED'' experiments we investigate the noise properties of coherently driven nonlinear resonators. 
By using Josephson junctions in superconducting circuits, strong nonlinearities 
can be engineered, which lead to the appearance of pronounced effects already 
for a low number of photons in the resonator. 
Based on a master equation approach we determine the emission spectrum and observe for typical circuit QED parameters, 
in addition to the primary Raman-type peaks, second-order peaks. 
These peaks describe higher harmonics 
in the slow noise-induced  fluctuations of the oscillation amplitude of  the resonator and provide a clear signature of the nonlinear nature of the system. 
\end{abstract}


\maketitle

\section{Introduction}

Several recent circuit QED experiments based on Josephson junctions in superconducting circuits were concerned with the 
properties of driven nonlinear resonators~\cite{Siddiqi2006}.  Josephson junctions allow engineering strong 
nonlinearities, which can lead to pronounced quantum effects, such as non-classical photon number distributions
 \cite{Marthaler2008}. 
At the same time, without leaving the quantum regime, the damping of the resonator can be made sufficiently strong to measure the 
radiation emitted by the resonator. State-of-the-art measurements give access to the spectral density of the emitted radiation 
$S(\omega)$. This yields detailed information about the quantum and classical fluctuations of the resonator 
\cite{CarmichaelBook,Drummond1981,Dykman2011}.
%, which is an interesting and challenging problem,
% since we are dealing with a nonlinear system out of equilibrium \cite{Andre2010}. 

Some of the experiments with coherently driven resonators based on a Josephson junction 
%embedded in a superconducting transmission line resonator 
were concerned with the development of a Josephson bifurcation amplifier
 to be used as a high-contrast readout device for superconducting qubits \cite{Metcalfe2007}. 
The Josephson bifurcation amplifier makes use of the dynamical bistability induced by a linear periodic 
driving with frequency $\omega_F$ close to the oscillator eigenfrequency
$\omega_0$.
%  which
% leads to a coexistence of two stable states with low- and high-amplitude oscillations, respectively. 
In the readout process, the two qubit states are mapped onto the two stable vibrational states of the resonator, 
which can be easily distinguished,
 since they differ strongly in amplitude and phase. The measurment backaction depends on the oscillator power
 spectrum 
 \cite{Serban2010,Clerk2010}.

The stable vibrational states are described by  nearly sinusoidal oscillations with frequency
 $\omega_F$. They are separated by a dynamical barrier in phase space \cite{Dykman1988a}.
 The amplitude and phase of the forced 
oscillations display quantum and thermal fluctuations. For typical parameters, the timescales for these fluctuations are set by the detuning 
$\delta\omega = \omega_F - \omega_0$ between the coherent driving and the resonator and by the relaxation rate $\Gamma$. %In this regime, the dynamical barrier can be thought as a quasi-energy Hamiltonian surface, which becomes time independent in a phase space rotating with frequency $\omega_F$ .

%In the bistable regime, fluctuations are small on average
%since for large quantum fluctuations \cite{PeanoDuffing,Vierheilig2010} 
%the oscillator density matrix is smeared over the basin of attraction of both classically stable states.\

Here we consider the regime, where the fluctuations are small on average and the oscillator stays in the 
vicinity of one of the stable solutions for times much longer than the fluctuations characteristic timescales,  
$\Gamma^{-1}$ and  $\left| \delta\omega \right|^{-1}$.  
Many theoretical investigations have focused on rare large fluctuations which allow overcoming the dynamical barrier and induce the
 switching between the stable solutions \cite{Dykman1988a,Marthaler2006}
 or on
tunneling \cite{FinkelsteinTunneling1976}. The deep quantum regime, where the average number of excitations in the resonator
 is small and quantum fluctuations are large has also attracted considerable attention \cite{PeanoDuffing}. 
The power spectrum in this regime has been investigated in \cite{Leyton}.

Although the emission spectrum $S(\omega)$  in principle carries information about rare large fluctuations 
\cite{Dykman2011,Dykman2012}, most of the radiation is emitted as a consequence of small fluctuations. Therefore, the  
standard approach for computing the spectrum $S(\omega)$ is to 
linearize the motion around the stable solutions \cite{Serban2010,Drummond1980c,Dykman2011,Dykman2012}. 
The  small fluctuations display damped harmonic oscillations with frequency $\sim |\delta\omega|$. 
The decay of the fluctuation correlations is accompanied by the emission of radiation with frequencies $\omega_F-\delta\omega$ and $\omega_F+\delta\omega$, since the slow oscillations are superimposed to the fast oscillations with the driving frequency. The resulting emission spectrum is reminiscent of the Raman spectrum of a diatomic molecule: beside the main Rayleigh peak at frequency $\omega_F$, which describes radiation emitted coherently, it has two Raman-type side-peaks with Raman shift $\pm |\delta\omega|$. In a quantum mechanical picture, the quasi-energy level spacing, i.e., the distance between the eigenstates of the Hamiltonian in a frame rotating with the driving frequency, is approximatively given by $\hbar \left| \delta\omega \right|$, and the Raman lines are induced by transition between nearest-neighbor quasi-energy eigenstates. In this context, the same driving field  defines and probes the  quasi-energy spectrum.

In this paper, we are interested in a regime where typical fluctuations are intermediate in size: 
they are too weak to induce switching between the two solutions, so that the resonator stays locked to one of the stable solutions for a long time, but they are strong 
enough to lead to nonlinear dynamics, so that one has to go beyond the simple Raman picture for the emission spectrum. 
It has been shown that quantum fluctuations of intermediate size may give rise to a fine structure in the spectrum \cite{Dykman2011,Dykman2012}. 

In this work we point to two additional features of the emission spectrum, which are consequences of the nonlinear nature of the quasi-energy Hamiltonian and should be observable in experiments \cite{WallraffSpectrumMeasurment}:
i) Intermediate strength nonlinear fluctuations are not simply damped sinusoidal oscillations but have higher harmonics yielding additional peaks in the resonator emission spectrum at frequencies $\omega_F\pm n\delta\omega$, $n=2,3,\cdots$. These higher order peaks resembles  the  peaks in the power spectrum
of a static  oscillator with a small cubic anharmonicity close to integer multiples of its eigenfrequency \cite{DykmanRussian}. At the quantum level, the higher harmonics derive from transitions between quasi-energy levels which are not nearest neighbors. ii) Moreover,  fluctuations in an asymmetric quasi-energy potential yield an additional broad peak at the frequency $\omega_F$. Unlike the fine structure of the emission spectrum, the effects we consider here are not restricted to the quantum regime but extend to the classical regime where thermal fluctuations are dominant. 

The paper is organized as follow: In Sec.~\ref{Sec:Model} we introduce the model for the driven nonlinear resonator and the approach based on a master equation. In Sec.~\ref{Sec:Smallfluct} we review the mean field and small fluctuations theory  for the oscillator. In Sec.~\ref{Sec:Spectrum} we outline our analytical calculation of the emission spectrum.  In Sec.~\ref{Sec:Numres} we compare our analytical finding with numerical results which are exact within our master equation approach.

\section{The model} \label{Sec:Model}

We consider a Duffing oscillator with  eigenfrequency $\omega_0$, coordinate $q$, 
momentum $p$, and Hamiltonian
\begin{equation}
H(t)=\frac{p^2}{2}+\frac{1}{2} \omega_0^2 q^2 + \frac{\gamma}{4} q^4 - F q \cos(\omega_F
t)\,.  \label{ham}
\end{equation}
             
We assume that the detuning $\delta\omega=\omega_F-\omega_0$ of the driving is small, $|\delta\omega|\ll \omega_0$, and that the nonlinearity for typical values of $q$
satisfies the condition $|\gamma|\langle q^2\rangle\ll \omega_0^2$. In this regime, the oscillator displays fast sinusoidal oscillations, $\langle q(t)\rangle \approx A(t)\cos [\omega_F t+\phi(t)]$, with amplitude $A(t)$ and phase $\phi(t)$ which vary slowly on the  timescale $\left| \delta\omega \right|^{-1}$.

It is convenient to study the oscillator dynamics in a frame rotating with the driving frequency.
I.e., we perform a unitary transformation $\tilde{H}=U^\dagger H U-i\hbar U^\dagger\dot{U}$ with $U(t)=\exp [-i\omega_F \hat{a}^\dagger \hat{a} \, t]$, where 
%$\hat{a}$ and $\hat{a}^\dagger$ are the ladder operators of the oscillator, 
$a=\left(\omega_0 q +i p\right)/\sqrt{2\hbar\omega_0}$. In the rotating frame, 
the ladder operators $a$ and $a^\dagger$ vary slowly, $\langle  a(t)\rangle\approx\sqrt{\omega_0 2 \hbar}A(t)\exp(-i\phi(t))$. We therefore use the rotating wave approximation (RWA) (neglecting fast oscillating terms with frequencies $2\omega_F$ and $4\omega_F$) and arrive at the time-independent Hamiltonian 
\begin{equation}
 \tilde{H}  \approx -\hbar \delta\omega \, a^{\dagger} a + \frac{\hbar K}{2}a^{\dagger} a(a^{\dagger} a+1)  -\frac {\hbar f}{2} \left( a + a^{\dagger} \right),
\label{eq:H-LabFrame}
\end{equation}
with $K=3\hbar\gamma/2\omega_0^2$ and $f=F/\sqrt{2\hbar\omega_0}$. 

We consider  Markovian dissipation due to a linear coupling  to a bosonic bath. When the oscillator relaxation rate is much smaller than its eigenfrequency, $\Gamma \ll \omega_0$, and the bath spectral density close to  $\omega_0$ is smooth, the dissipative dynamics of the system can be described by a simple master equation in the Lindblad form 
\begin{equation}
\dot{\rho} = {\cal L}\rho = -\frac{i}{\hbar}[\tilde{H},{\rho}]+
  \Gamma(1+\bar{n}){\cal D}[a]{\rho} +  \Gamma \bar{n}{\cal
 D}[a^{\dagger}]{\rho},  \label{eq:Liouville}
\end{equation}
where the Lindblad superoperator is defined through ${\cal D}[O]{\rho}\equiv 2O{\rho}O^{\dagger}
-O^{\dagger}O{\rho}-{\rho}OO^{\dagger}$,   and $\bar{n}=(e^{\hbar\omega_0/k_BT}-1)^{-1}$ is the oscillator
distribution in the absence of driving. 

For long times the resonator relaxes to a stationary state, satisfying ${\cal L} \rho^{\rm st} = 0$, 
from which we obtain mean values  $\langle O\rangle_{\rm st} \equiv \mathrm{Tr}\langle\rho^{\rm st}O\rangle$ in the rotating frame. In particular, it allows us to calculate the oscillation amplitude $A^{\rm st}=\sqrt{2\hbar/\omega_0} \langle a\rangle_{\rm st}$.
  
The Lindblad master equation (\ref{eq:Liouville}) gives also access to the full time-evolution of the system, and we can use it to calculate correlation functions $\langle O(t+\Delta t) O'(t) \rangle = \mathrm{Tr} \lbrace O e^{{\cal L}\Delta t} O' \rho(t) \rbrace$ \cite{Ginzel1993}, where ${\cal L}$ is the superoperator defined in 
Eq. (\ref{eq:Liouville}). For long times $t$,  the system is in the stationary state $\rho^{\rm st}$, so that the correlation function depends on the time difference $\Delta t$ only, $\langle O(t+\Delta t) O'(t) \rangle=\mathrm{Tr} \lbrace O e^{{\cal L} \Delta t} O' \rho^{\rm st} \rbrace\equiv\langle O(\Delta t) O' \rangle_{\rm st}$.

Here, we are specifically interested in the emission spectrum $S(\omega)$, i.e., the spectral density of the photons emitted by the driven resonator,
\begin{equation}\label{eq:emissionspectrum}
  S(\omega) = 2 \, \mathrm{Re} \int_0^{\infty} d  t \, \langle a^{\dagger} ( t) a \rangle_{\rm st} e^{-i (\omega-\omega_F)t}. 
\end{equation}
The above definition takes into account that the correlation function is computed in a frame rotating with frequency $\omega_F$.

The emission spectrum $S(\omega)$ consists of two distinct contributions coming from the finite mean value $\langle a \rangle_{\rm st}$ and from the fluctuations $\langle (a^{\dagger} ( t)-\langle a \rangle_{\rm st})( a-\langle a \rangle_{\rm st}) \rangle_{\rm st}$ of the operator $a$ (which describes the oscillation amplitude). The latter yield broad peaks with width of the order of the oscillator relaxation rate $\Gamma$ and height proportional to the  noise intensity. On the other hand, the finite mean value $\langle a \rangle_{\rm st}$ yields a sharp peak at the driving frequency $\omega_F$. For a strictly monochromatic driving, this peak has the shape of a delta function, $\pi \langle a\rangle_{\rm st}^2\delta(\omega-\omega_F)$. In the following, we subtract this contribution from the emission spectrum  $S(\omega)$ and focus  only on the broad peaks due to the fluctuations of $a$.

\begin{figure}
  \includegraphics[width=\columnwidth]{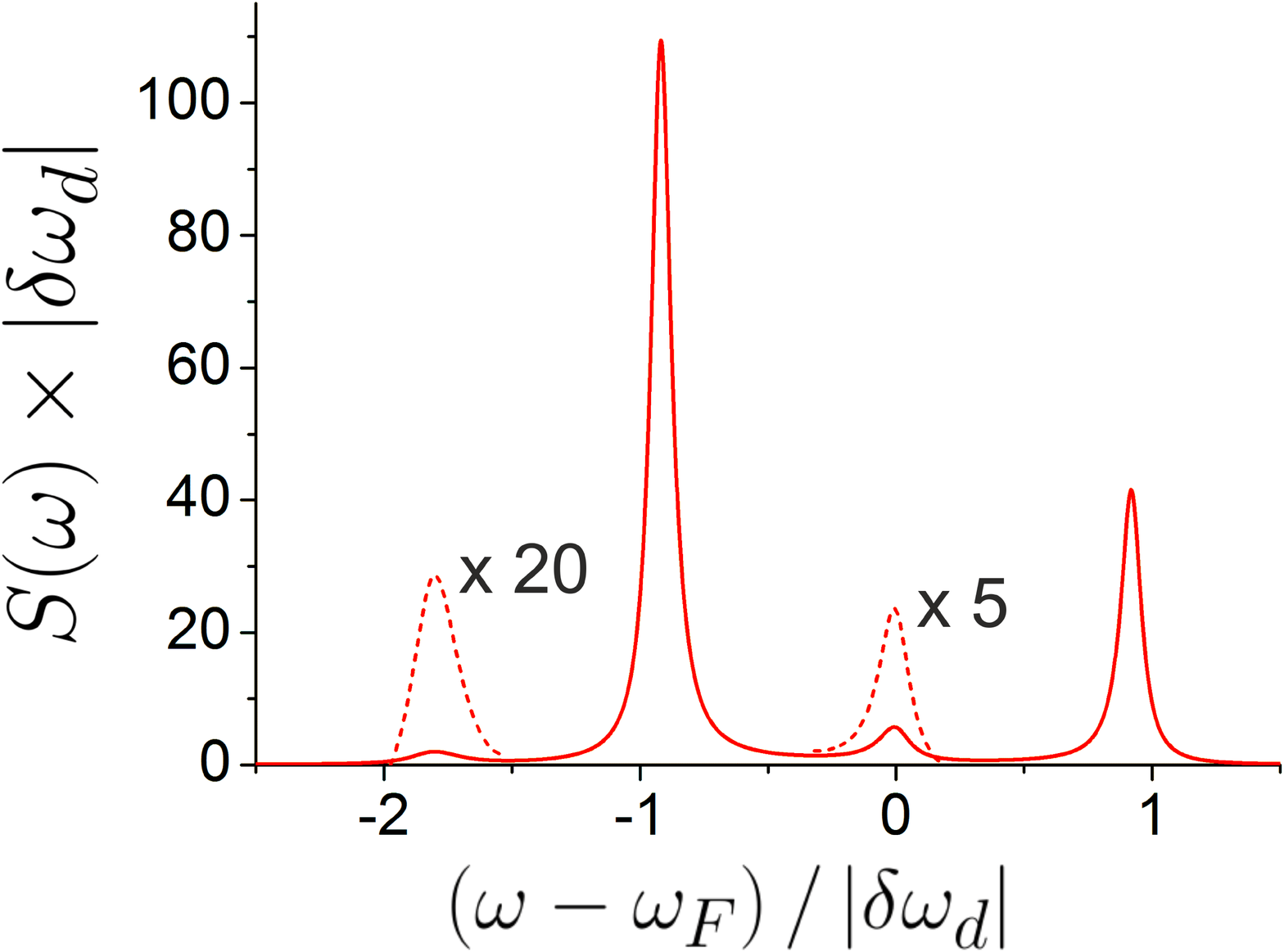}
\caption{Emission spectrum of the driven nonlinear resonator in the
 regime of high-amplitude oscillations. The parameters are: $\delta\omega=-75\textrm{MHz}$, $K=-3\textrm{MHz}$, $f=270\textrm{MHz}$, $\Gamma=3\textrm{MHz}$ and $\bar{n}=1$. The dashed lines show the second-order peaks 
scaled up by a constant factor.}
\label{fig:SpecIntro}
\end{figure}

In Fig.~\ref{fig:SpecIntro}, we show the emission spectrum $S(\omega)$ for a parameter choice in the  regime of operation of a Josephson bifurcation amplifier \cite{Siddiqi2006}. In this example, the lineshape of $S(\omega)$ consists of four distinct Lorentzian peaks: two high peaks of width  $\Gamma$
for   $\omega-\omega_F$ close to $\pm\delta\omega$ and two lower peaks of width $2\Gamma$ for $\omega-\omega_F$ close to $-2\delta\omega$ and $0$.  Such peaks are consistent with weakly damped oscillations of the fluctuation of $a$ with period $\delta\omega$, which are mostly harmonic but have  substantial second harmonic and  average over a period $2\pi/\delta\omega$. A linearized theory of the fluctuations around the mean-field solutions \cite{Drummond1980c,Dykman2011} yields damped harmonic oscillations for the fluctuations of $a$, but does not capture the second order peaks for $\omega-\omega_F\approx -2\delta\omega, 0$.
In order to go beyond this limitation  we will compute in Section \ref{Sec:Spectrum} the emission spectrum by taking into account the nonlinear nature of the fluctuations.

\section{The mean field solutions and small fluctuations}
\label{Sec:Smallfluct}

In this section, in order to prepare for the calculation of the emission spectrum, we review
the mean field  and small fluctuation theories for the Duffing oscillator \cite{Drummond1980c,Dykman1988a,Dykman2012}. 

\subsection{Dynamical bistability}

For concreteness, we focus on soft nonlinearities, $K < 0 $ (which apply to  the Josephson nonlinearity), and red detuned driving, $\delta \omega<0$. We switch to dimensionless variables by introducing the time $\tau\equiv|\delta\omega| t$, the effective Planck constant 
\begin{equation}\label{eq:effPlanck}
\lambda\equiv\frac{K}{2\delta\omega}=\frac{3\hbar\gamma}{4\omega_0^2\delta\omega}\,,
\end{equation}
and the slowly varying quadratures $Q$ and $P$, 
\begin{equation}\label{slowquad}
 Q\equiv\sqrt{\frac{\lambda}{2}}(a+a^\dagger),\quad P\equiv\sqrt{\frac{\lambda}{2}}i(a^\dagger-a)\,.
\end{equation}
 with canonical commutator $[Q,P]=i\lambda$.
%\begin{eqnarray}
% U^\dagger q U &=&\sqrt{\frac{\hbar}{\lambda\omega_F}}\left[
%Q\cos \omega_F t-
%P\sin \omega_F t\right]\\
%U^\dagger p U & =& \sqrt{\frac{\hbar\omega_F}{\lambda}}\left[
%Q\sin \omega_F t+
%P\cos\omega_F t\right]\,.
%\end{eqnarray}
The dimensionless Schr\"odinger equation reads
\begin{equation}
 i\lambda\partial_\tau \psi(Q,\tau)=g(Q,P=-i\lambda\partial_Q)\psi(Q,\tau)\,.
\end{equation}
where the quasi-energy Hamiltonian
\begin{equation}\label{Hq}
g= -(Q^2+P^2-1)^2/4+\sqrt{\beta}Q.
\end{equation} 
depends on  the parameter $\beta= f^2 K/(4|\delta\omega|^3)$ only and is time-independent.
However, it can not be written as a sum of kinetic and potential energy. For hard nonlinearity, $K>0$,
and blue detuning, $\delta \omega>0$, the quasienergy surface $\hat{g}$  has  opposite sign \cite{Dykman1988a}.

\begin{figure}
  \includegraphics[width=\columnwidth]{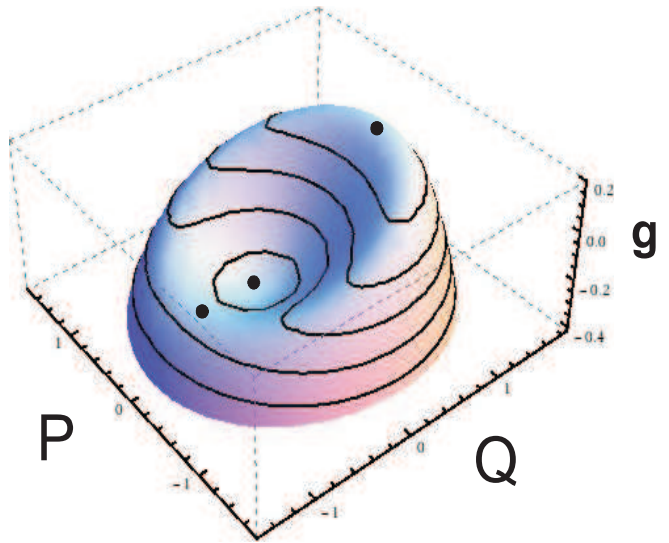}
  \caption{3D-plot of the function $g(Q,P)$ for a fixed value of the scaled driving $\beta = 0.034$. The 
   minimum corresponds to the small amplitude state and the maximum corresponds to the high amplitude state.}
  \label{Fig:quasi-energy3D}
\end{figure}

The values of quasi-energy Hamiltonian $g(Q,P)$ as function of the variables $Q,P$ are shown in the 3D plot in  Fig.~\ref{Fig:quasi-energy3D} for a fixed value of $\beta=0.034$. It has three stationary points corresponding to a minimum, a maximum, and a saddle point. These stationary points coexist for $0<\beta<4/27$. The maximum and the minimum are located at 
\begin{equation}\label{large}
(Q,P)=(Q_{\rm h}\equiv\cos\theta/\sqrt{3}+\sin\theta,P_{\rm h}\equiv0)
\end{equation}
and
\begin{equation}\label{small}
(Q,P)=(Q_{\rm l}\equiv\cos\theta/\sqrt{3}-\sin\theta,P_{\rm l}\equiv0),                                                       
\end{equation}
respectively.

Here, the angle $\theta$ is given by $\theta=(\pi-\arctan\sqrt{4/(27 \beta)-1})/3$. 
In the laboratory frame, these two solutions  describe  sinusoidal oscillations with frequency $\omega_F$, 
dimensionless oscillation amplitude  $|Q_{\rm l}|$ and $|Q_{\rm h}|$, with $|Q_{\rm l}|<|Q_{\rm h}|$, and opposite phases,  $\phi_{\rm l}=0$ and $\phi_{\rm h}=\pi$. 
They describe the low- and high-amplitude solutions, respectively.

For  weak  damping, $\kappa\equiv\Gamma/ \left| \delta\omega \right| \ll1$, the low- and high-amplitude states become attractors. For weak quantum fluctuations, $\lambda\ll 1$,
there is a separation of time scales: On the timescale $\sim\kappa^{-1}$, the oscillator relaxes to the vicinity of either one of the two states, while on a much 
longer timescale $\propto\exp[\lambda^{-1}]$ rare large thermal or quantum fluctuations induce switching between them. In the next section, 
we focus on the quasi-stationary regime,  $\kappa^{-1}\ll\tau\ll\kappa^{-1}\exp[\lambda^{-1}]$, where the oscillator stays in the vicinity of 
one of the two metastable solutions.

\subsection{Small fluctuations around the mean field solutions}
 
In the weak damping regime, $\kappa\ll 1$, small quantum  and thermal fluctuations 
around a mean field solution can be taken into account in two steps. In the first one introduces
 an auxiliary  oscillator, which describes the quantum states localized in the phase-space region close to one of the two solutions. In the second one introduces an effective master equation for the auxiliary oscillator. It turns out that the fluctuations and the dissipation for the auxiliary oscillator are similar as for an oscillator close to equilibrium but with an effective temperature $T_e$, to be discussed below \cite{Dykman2011}.

The auxiliary oscillator is obtained by expanding the quasi-energy Hamiltonian
 (\ref{Hq}) around the classical solutions $Q_\mathrm{i}=Q_{\rm l/h}$ and $P=0$, where the index $i$ denotes the states of low- (i=l) and high- (i=h) amplitude oscillations:
\begin{equation}
 \hat{g}\approx g_\mathrm{i} + \frac{1}{2}g_{PP}P^2+\frac{1}{2}g_{QQ} (Q-Q_\mathrm{i})^2
\end{equation}
with $g_\mathrm{i}=g(Q_\mathrm{i},0)$, $g_{PP}=Q_\mathrm{i}^2-1$ and $g_{QQ}=3Q_\mathrm{i}^2-1$.

It is convenient to rewrite the Hamiltonian in terms of ladder operators 
\begin{equation}\label{eq:linham}
 \hat{g}\approx g( Q_\mathrm{i},0)+\mathrm{sgn}(g_{QQ})\lambda\, \nu_\mathrm{i}(b^\dagger b+1/2)\,
\end{equation}
where  $\nu_\mathrm{i}=\sqrt{g_{QQ}g_{PP}}$ is the dimensionless frequency of the slow oscillations and the ladder operators $b$ and $b^\dagger$ are obtained by the squeezing transformation 
\cite{Walls2008}
\begin{equation}\label{eq:sqtrans}
 a=a_\mathrm{i} + b \cosh r_\mathrm{i}^*-b^\dagger\sinh r_\mathrm{i}^*
\end{equation} 
with $a_\mathrm{i} = Q_\mathrm{i}/\sqrt{2\lambda}$ and  squeezing factor $r_\mathrm{i}^*=\ln[g_{QQ}/g_{PP}]/4$. 

For $\kappa\ll\nu_0$ we can incorporate small thermal fluctuations by substituting the squeezing transformation Eq.\ (\ref{eq:sqtrans}) into the Lindblad master equation (\ref{eq:Liouville}) and neglecting the terms where the operators $b$ and $b^\dagger$ are not 
matched in pairs. We find
\begin{equation}\label{efME}
\partial_\tau \rho=-\frac{i}{\lambda}[g,{\rho}]
 + \kappa(1+\bar{n}_e){\cal D}[b]{\rho} +  \kappa \bar{n}_e{\cal
 D}[b^{\dagger}]{\rho}, 
\end{equation}
with $\bar{n}_e=\bar{n}+(2\bar{n}+1)\sinh^2 r_\mathrm{i}^*$~\cite{Dykman2011}. This  master equation has the same structure as the master equation for a static  weakly-nonlinear oscillator with eigenfrequency $\nu_\mathrm{i}$ and damping $\kappa$ at the temperature $T_{\rm e}=\lambda\nu_\mathrm{i}/\ln[(\bar{n}_e+1)/\bar{n}_e]$. It naturally yields a stationary distribution for the auxiliary oscillator Fock states  $|n\rangle$ of the Boltzmann 
form \cite{Dykman2011,Dykman2012}
\begin{equation}
 \rho_n^{\rm st}\equiv\langle n|\rho^{\rm st}|n\rangle\approx (\bar{n}_e+1)^{-1}\exp[-n \lambda\nu_\mathrm{i}/T_e]\,.
\end{equation}
Note that for a driven oscillator, the effective temperature $T_{\rm e}$
 is finite even for $\bar{n}=0$, $T_{\rm e}=\lambda\nu_\mathrm{i}/(2\ln \coth r_\mathrm{i}^*)$~\cite{Peano2010,Dykman2011,Dykman2012}, 
a phenomenon which has been named quantum heating \cite{Dykman2011}. In the classical limit of high temperatures $k_B T\gg \hbar\omega_0$, we have \cite{Dykman2011}
\begin{equation}\label{eq:T_eclass}
T_e\approx\lambda\nu_\mathrm{i}\bar{n}_e\approx k_B T \frac{3 \gamma}{4 \omega_0^3 \delta \omega}(1+2\sinh^2 r_\mathrm{i}^*)\, ,
\end{equation}
so that $T_e$  becomes independent of $\hbar$, as it should be expected for a classical quantity.

\section{The emission spectrum} \label{Sec:Spectrum}

The master equation for the auxiliary oscillator Eq.\ (\ref{efME}) allows us to compute the emission spectrum $S(\omega)$ close to the Lorentzian peaks, in the limit where they are well separated, $\kappa\ll\nu_\mathrm{i}$. In general, it is not sufficient to use the  harmonic oscillator approximation for the auxiliary oscillator, rather one has to include  in Eq.\ (\ref{efME}) the relevant nonlinear corrections to $\hat{g}$. To proceed, we focus on the parameter regime  where the second order peaks are clearly visible. In this case, we can neglect in the quasi-energy Hamiltonian $\hat{g}$ (\ref{Hq}) terms which are quartic in  the ladder operators $b$ and $b^\dagger$. Thus we find
\begin{equation}\label{eq:cubicquasien}
 g\approx g_\mathrm{i}+ \mathrm{sgn}(g_{QQ})\lambda \nu_\mathrm{i} b^\dagger b+ \lambda^{3/2}\left(V_1 b^3+V_2 b^{\dagger 2} b  +    
 h. c.\right)\,,                                                       
\end{equation}
with  $V_1=Q_\mathrm{i}( e^{-3 r_\mathrm{i}^*}-e^{r_\mathrm{i}^*}) /2^{3/2}$ and $V_2=Q_\mathrm{i} (3e^{-3 r_\mathrm{i}^*}+e^{ r_\mathrm{i}^*})/2^{3/2}$.  

In order to compute the emission spectrum $S(\omega)$, we insert the squeezing transformation 
Eq.~(\ref{eq:sqtrans}) into the  definition (\ref{eq:emissionspectrum}) of the emission spectrum 
and obtain the expression 
\begin{eqnarray}\label{powerbbdag}
S(\omega)&=& \frac{1}{\left| \delta\omega \right|}\left(\cosh^2 r_\mathrm{i}^* 2{\rm Re}\langle b^\dagger, b\rangle_\nu+
\sinh^2 r_\mathrm{i}^*2{\rm Re}\langle b, b^\dagger\rangle_{\nu}
\right.\nonumber\\&&\left.
-2\sinh r_\mathrm{i}^*\cosh r_\mathrm{i}^*2{\rm Re}\langle b, b\rangle_{\nu}\right)\,.
\end{eqnarray}
Here, we have defined 
\begin{equation}
\nu=(\omega-\omega_F)/|\delta\omega| 
\end{equation}
and the spectral functions
\begin{equation}
\langle O, O'\rangle_\nu\equiv \int_{0}^\infty d\tau e^{-i\nu \tau}\langle O(\tau) O'\rangle_{\rm st}\,,
\end{equation}
The evaluation of the auxiliary oscillator spectra $\langle b, b^\dagger\rangle_\nu$, $\langle b^\dagger, b\rangle_\nu$ and $\langle b, b\rangle_\nu$
proceeds along the same lines as the calculation of the spectrum of a weakly-nonlinear oscillator
 \cite{DykmanRussian,Dykman1988}, and is detailed in Appendix \ref{App:PowerSp}. 
By substituting Eqs.\ (\ref{eq:bbd}), (\ref{eq:bdb}) and (\ref{eq:bb}) into Eq.\ (\ref{powerbbdag}) 
we obtain the results for 
$S(\omega)$ close to its peak.
% \begin{equation}
% S(\omega)=\sum_{n=-2}^2 \frac{S_n\kappa_n
% }{(\nu-n\,\mathrm{sgn}(g_{QQ})\nu_0)^2+\kappa_n^2}
%\end{equation}
%$S(\omega)=\sum_{n=-2}^2 S_n(\omega)$ where

For frequencies $\nu$ close to $\nu_\mathrm{i}$ and $-\nu_\mathrm{i}$ we find the first order peaks 
\begin{equation}
S(\omega)= \frac{1}{\left| \delta\omega \right|} \frac{2\kappa\cosh^2 r_\mathrm{i}^*
 \bar{n}_e}{(\nu-\mathrm{sgn}(g_{QQ})\nu_\mathrm{i})^2+\kappa^2}\,,
\label{eq:Spectrum1P}
\end{equation}
and
\begin{equation}
 S(\omega)= \frac{1}{\left| \delta\omega \right|} \frac{2\kappa\sinh^2 r_\mathrm{i}^*(\bar{n}_e+1)
 }{(\nu+\mathrm{sgn}(g_{QQ})\nu_\mathrm{i})^2+\kappa^2}\,,
\label{eq:Spectrum1M}
\end{equation}
respectively. The expressions for the first order peaks have been obtained previously using a quantum Fokker-Planck equation approach \cite{Drummond1980c}. 

For frequencies $\nu$ close to zero we find
\begin{equation}
 S(\omega)= \frac{1}{\left| \delta\omega \right|}\frac{\lambda Q_\mathrm{i}^2}{\nu_\mathrm{i}^2}(3e^{-4r_\mathrm{i}^*}+1)^2 \frac{2\kappa\bar{n}_e(\bar{n}_e+1)}{\nu^2+4\kappa^2}\,. 
 \label{eq:Spectrum0} 
\end{equation}
The central peak is a consequence of amplitude fluctuations $\delta a_n\equiv\langle n|a|n\rangle-\langle a\rangle_{\rm st}\neq 0$ of the quasi-energy states. It is in fact possible to rewrite eq. (\ref{eq:Spectrum0}) as $2 (d\delta a_n/dn)^2  \langle b^\dagger b,b^\dagger b\rangle_\nu$ \cite{Dykman1988}. The peak is therefore determined by a combination of the amplitude curvature and quasi-energy fluctuations. 
Close to $\pm 2\nu_0$ we find the second-order peaks
\begin{equation}
  S(\omega)=\frac{1}{\left| \delta\omega \right|} \frac{\lambda Q_\mathrm{i}^2}{\nu_\mathrm{i}^2e^{4 r_\mathrm{i}^*}} \frac{16\kappa \cosh^4 r_\mathrm{i}^*\bar{n}_e^2}{(\nu-\mathrm{sgn}(g_{QQ})2\nu_\mathrm{i})^2+4\kappa^2} 
  \label{eq:Spectrum2P}
\end{equation}
\begin{equation}
  S(\omega)=\frac{1}{\left| \delta\omega \right|} \frac{\lambda Q_\mathrm{i}^2}{\nu_\mathrm{i}^2e^{4 r_\mathrm{i}^*}}\frac{16 \kappa\sinh^4 r_\mathrm{i}^*(\bar{n}_e+1)^2}{(\nu+\mathrm{sgn}(g_{QQ})2\nu_\mathrm{i})^2+4\kappa^2} 
 \label{eq:Spectrum2M}  
\end{equation}
Equations (\ref{eq:Spectrum0}), (\ref{eq:Spectrum2P}) and (\ref{eq:Spectrum2M}) 
are valid only close to the corresponding peaks, which  do not have a substantial overlap for $\kappa^2\ll\lambda$. Moreover they do not incorporate
a shift in the peak center of order $\lambda \left| \delta\omega \right|$, and do not take into
account the non equidistance of the quasienergy levels, which give rise to a hyperfine splitting similar to Refs. \onlinecite{Dykman2011,DykmanRussian}, when it 
exceeds the levels broadening, for $\kappa\ll\lambda$. Therefore strictly speking they are valid for $\kappa^2\ll\lambda\ll\kappa$.

In the classical  limit $k_BT\gg\hbar\omega_0$, the intensity of the emitted power irradiated $\Gamma \hbar \omega S(\omega)$ becomes independent of  $\hbar$ as can be verified by substituting   Eqs. (\ref{eq:effPlanck},\ref{eq:T_eclass}) into the lineshape Eqs. (\ref{eq:Spectrum1P}
-\ref{eq:Spectrum2M}) of $S(\omega)$ close to its peaks.

\begin{figure}
  \includegraphics[width=\columnwidth]{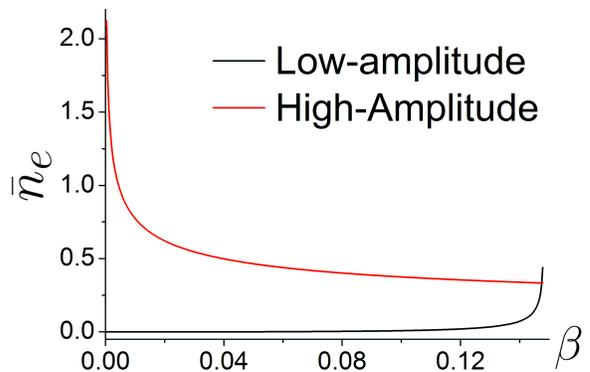}
  \caption{The quantum temperature $\bar{n}_e$  for the two stable states as function of the scaled driving strength $\beta$ for $\bar{n}=0$
 (see  Ref. \onlinecite{Dykman2012}).}  
  \label{Fig:QuantumTemp}
\end{figure}
The  amplitude of the first order peaks is linear in the quantum temperature $\bar{n}_e$, while the second order peak are proportional to the squared noise intensity $\bar{n}_e^2$. Fig. \ref{Fig:QuantumTemp} shows the quantum temperature $\bar{n}_e$ of the two stable states for $\bar{n} =0$. As we can see, in the state of low-amplitude oscillations, the quantum temperature is mostly rather small. It becomes larger for high values of the scaled driving strength $\beta$; however, in this regime, the stationary state of the driven resonator is the state of high-amplitude oscillations. On the other hand, for high-amplitude oscillations, the quantum temperature takes large values.

\section{Results}
\label{Sec:Numres}

In this section we present the results for the emission power spectrum. We mostly show results from a numerical solution of Eq.\ (\ref{eq:Liouville}) with boundary condition $\rho(0)=a\rho_s$, which  also covers parameters beyond the regime $\kappa^2 \ll \lambda \ll \kappa$, where the analytic solution is valid.

\begin{figure}
  \includegraphics[width=\columnwidth]{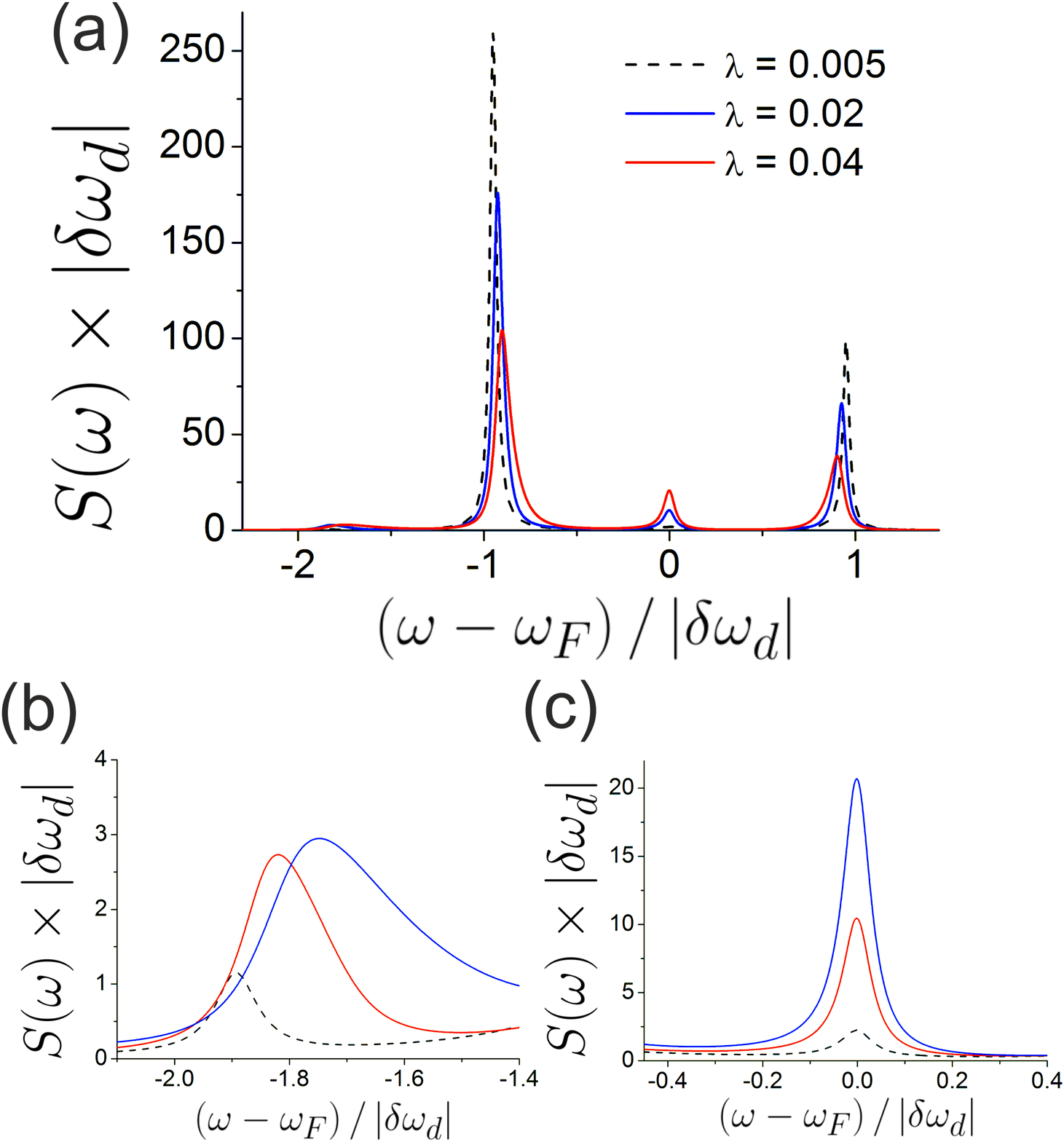}
  \caption{(a) Emission spectrum of the driven nonlinear resonator in the regime of low-amplitude  oscillations for three different values of the parameter $\lambda$. 
The other parameters are $\beta = 0.13$, $\kappa = 0.02$ and $\bar{n} = 1$.
(b) and (c) shows a zoom around the second order peaks at the frequencies $\omega \approx -2 \left| \delta \omega \right|$ and $\omega = 0$. }
\label{fig:SpecHA}
\end{figure}

First we focus on the case of strong driving, where the resonator is in a stationary state of high-amplitude oscillations. In Fig.~\ref{fig:SpecHA} we show the emission spectrum for finite temperature, $\bar{n} = 1.0$ and $\beta = 0.13$. In the regime of high-amplitude oscillations, the squeezing factor $r_\mathrm{h}^*$ is rather large, so that according to Eqs. (\ref{eq:Spectrum1P}) and (\ref{eq:Spectrum1M}), both first-order peaks are of comparable size. 

We observe the appearance of two second-order peaks. The first one is located at the frequency $\omega \approx - 2 \left| \delta \omega \right|$ and is described by Eq. (\ref{eq:Spectrum2P}). The oscillation amplitude $Q_\mathrm{h}$ is large, $\left| Q_\mathrm{h} \right| \approx 1$, so that the peak height $\sim \lambda / \kappa$ is larger than the background $\sim \kappa$ arising from the first-order peaks. We also observe  a second-order peak at the frequency $\omega = 0$, which is described by Eq. (\ref{eq:Spectrum0}). 
We note that for $\lambda = 0.005$, where the condition $\kappa^2 \ll \lambda \ll \kappa$ is fulfilled, the numerical results are very well described by the analytical expressions (\ref{eq:Spectrum1P})-(\ref{eq:Spectrum2M}). 

The analytical expressions we derived for the emission spectrum predict that the height of the second-order peaks is proportional to the ratio $\lambda / \kappa$. Therefore the second-order peaks become more visible against the  background for larger values of the parameter $\lambda$. This is also shown in Fig.~\ref{fig:SpecHA}, where we plot the emission spectrum for different values of the parameter $\lambda$. 

We see that, while the first-order-peaks are decreasing, both second-order peaks are clearly increasing for larger values of $\lambda$, consistent with the analytical predictions.For the large values of $\lambda$ used in the plot, we are close to  the limit  of validity of the analytical expressions for the spectrum, since the condition $\lambda \ll \kappa$ is not strictly fulfilled. In this regime our analytical result still provide  the correct qualitative picture for the emission spectrum and a remarkably good estimate of the line intensities, but it does not describe a finite shift of the peaks positions. 

%%%%%%%%%%%%%%%%%%%%%%%%%%%%%%%%%%%%%%%%%%%%5

\begin{figure}
  \includegraphics[width=\columnwidth]{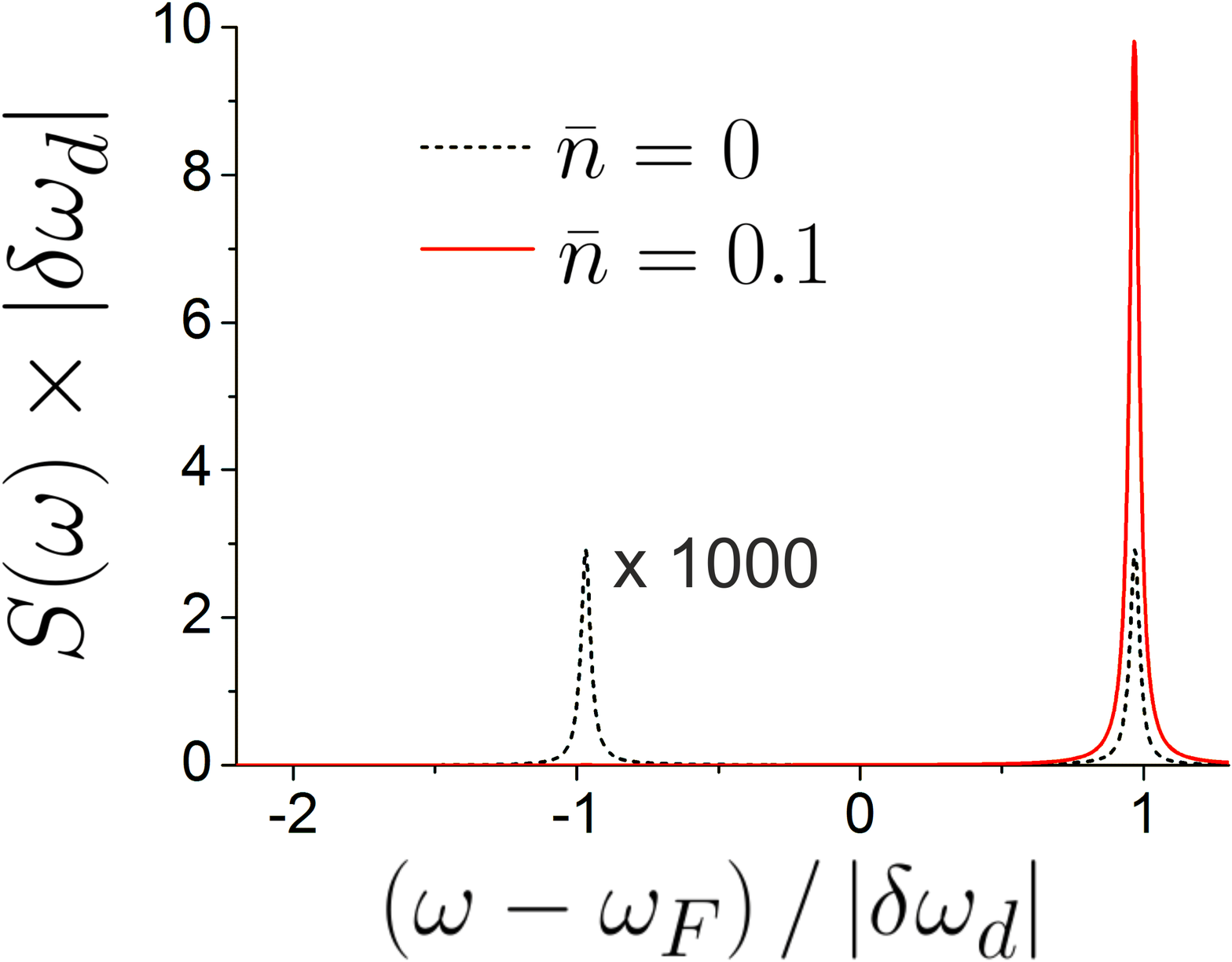}
  \caption{Emission spectrum of the driven nonlinear resonator in the regime of low-amplitude oscillations. The parameters are $\beta = 0.01$, $\lambda = 0.005$ and $\kappa = 0.02$.} 
  \label{fig:SpecLA}
\end{figure}

For the sake of comparison, we also discuss the case of weak driving , where the resonator is in a state of low-amplitude oscillations. The emission spectrum is shown in Fig.~\ref{fig:SpecLA} for zero and finite temperature, and for fixed driving, $\beta = 0.01$. Here only the first-order peaks, located at the frequencies $\omega \approx \pm \left| \delta \omega \right|$ are visible. 

Since the oscillation amplitude $Q_\mathrm{l}$ is small, $Q_\mathrm{l} \ll 1$, the squeezing factor $r_\mathrm{l}^*$ is close to zero, $r_\mathrm{l}^* \approx 0$, and the effective temperature is close to the physical temperature, $\bar{n}_e \approx \bar{n}$. This means that for finite temperature, $\bar{n} > 0$, the first-order peak $\sim \bar{n}_e \cosh^2 r_\mathrm{l}^*$ at the frequency $\omega \approx \left| \delta \omega \right|$ is much larger than the second first-order peak $\sim \left( \bar{n}_e + 1 \right) \sinh^2 r_\mathrm{l}^*$ at the frequency $\omega \approx \left| \delta \omega \right|$.

On the other hand, in the limit of zero temperature, when $\bar{n}_e$ tends to zero, both first-order peaks are of the same (small) size. Finally, the second-order peaks $\sim \lambda Q_0^2$ are so small compared to the background of the first-order peaks that they are not visible. \\

\section{Summary}

We have studied the emission spectrum of a driven nonlinear oscillator. We focused on a regime characterized by a strong nonlinearity and low temperature, 
which are typical for circuit QED experiments and lead to the appearance of pronounced nonlinear effects.

In the case that the oscillator is locked to the regime of high-amplitude oscillations, we found
two new features in the spectrum, in addition to the Raman-side peaks located at frequencies $\omega\sim\pm \left| \delta \omega \right|$: 
i) We observed the appearance of a second-order peak at the frequency $\omega\sim\omega_F - 2 \left| \delta \omega \right|$. 
In a quantum-mechanical picture, this emission peak results from transitions between next-to nearest quasi-energy levels. ii) We observed a second-order peak at the 
driving frequency $\omega_F$ (in addition to the sharp Rayleigh peak) which results from fluctuations in the asymmetric quasi-energy potential.

Based on a master equation approach, we derived analytical expressions for the emission spectrum around the first- and second- order emission peaks.
 The results were obtained by expanding the quasi-energy Hamiltonian around the classical stable solutions. In order to describe the second-order peaks,
 we went beyond the usual linearization of the equations of motion and accounted for the nonlinear nature of the quasi-energy Hamiltonian. 

\section*{Acknowledgements}

We thank Mark Dykman, Xin-Qi Li, Vicente Leyton Ortega, Alexander Shnirman and Michael Thorwart  for stimulating discussions. 
The research of VP was supported by
the NSF, grant EMT/QIS 082985.

\appendix
\section{Calculation of the spectra for the auxiliary nonlinear oscillator}\label{App:PowerSp}
In this Appendix we compute the spectra $\langle b, b^\dagger\rangle_\nu$,  $\langle b^\dagger, b\rangle_\nu$ and $\langle b, b\rangle_\nu$,
for the auxiliary nonlinear oscillator, defined in  Eq.\ (\ref{eq:cubicquasien}),
whose dissipative dynamics is governed by the Lindblad master equation (\ref{efME}). We start with
\begin{equation}
 \langle b, b^\dagger\rangle_\nu\equiv \int_0^\infty d\tau\, e^{-i\nu \tau}\langle b(\tau) b^\dagger\rangle_{\rm st}
\end{equation}
 By integrating by parts we get
\begin{equation}\label{eq:partder}
 \langle b, b^\dagger\rangle_\nu
=-\frac{i}{\nu}\langle {\dot b}, b^\dagger\rangle_\nu-\frac{i}{\nu}\langle  b b^\dagger\rangle\,.
\end{equation}
The equation of motion   for  $\langle b\rangle$,
$\langle \dot{b}\rangle=i/\lambda\langle[g,b]\rangle-\kappa \langle b\rangle$  \cite{Walls2008},  
directly applies to $\langle b(t)b\rangle_{\rm st}$, since $\langle b(t)b\rangle_{\rm st}=\langle b(t)\rangle$ 
with initial condition $\langle b(0)\rangle=\mathrm{tr}\langle b^\dagger b\rho^{\rm st}\rangle$.
%\begin{equation}\label{eq:eqmotb}
% \langle \dot{b}\rangle=\frac{i}{\lambda}\langle[g,b]\rangle-\kappa \langle b\rangle=(-i\,\mathrm{sqn}(g_{QQ})\nu_0-\kappa) \langle b\rangle-i \langle B\rangle
%\end{equation}
%with $B=\lambda^{1/2}[3 V_1 b^{\dagger 2}+2V_2 b^\dagger b+V_2 b^{ 2}]$.
%Since $\langle b(t)b\rangle_{\rm st}=\langle b(t)\rangle$ with initial condition $\langle b(0)\rangle=\mathrm{tr}\langle b^\dagger b\rho^{\rm st}\rangle$,
% the equation of motion in Eq.\ (\ref{eq:eqmotb}) can be directly applied to the right hand side of Eq.\ (\ref{eq:partder}) yielding
%\begin{eqnarray}
% &&\langle \dot{b}(t)b\rangle_{\rm st}=\frac{i}{\lambda}\langle[g,b](t)b\rangle_{\rm st}-\kappa \langle b(t)b\rangle_{\rm st}\nonumber\\
%&=&(-i\,\mathrm{sqn}(g_{QQ})\nu_0-\kappa) \langle b(t)b^{\dagger}\rangle_{\rm st}-i \langle B(t)b^{\dagger}\rangle_{\rm st}
%\end{eqnarray}
% with
%$B=\lambda^{1/2}[3 V_1 b^{\dagger 2}+2V_2 b^\dagger b+V_2 b^{ 2}]$.
By inserting it, together with the approximate expression of the quasi-energy $\hat{g}$ into Eq.\  (\ref{eq:cubicquasien}), 
into Eq. (\ref{eq:partder}),  we find
\begin{eqnarray}\label{eq:bbd->Bbd}
 \langle b, b^\dagger\rangle_\nu
&=&-\frac{\langle B, b^\dagger\rangle_\nu+i\langle  b b^\dagger\rangle}{\nu+\mathrm{sgn}(g_{QQ})\nu_0-i\kappa}
\end{eqnarray}
with $B=\lambda^{1/2}[3 V_1 b^{\dagger 2}+2V_2 b^\dagger b+V_2 b^{ 2}]$.
Taking into account that
\begin{equation}
 \langle B, b\rangle_\nu\equiv \int_0^\infty d\tau e^{-i\nu \tau}\langle B(\tau) b^\dagger\rangle_{\rm st}=
\int_{-\infty}^0 d\tau e^{i\nu \tau}\langle B b^\dagger(\tau)\rangle_{\rm st}
\end{equation}
we can  integrate by parts and apply the equation of motion for $\langle b^\dagger\rangle$, obtaining
\begin{equation}
 \langle B, b^\dagger\rangle_\nu
=-\frac{\langle B, B^\dagger\rangle_\nu+i\langle  B b^\dagger \rangle}{\nu+\mathrm{sgn}(g_{QQ})\nu_0+i\kappa}\,.
\end{equation}
By substituting into Eq.\ (\ref{eq:bbd->Bbd}), we arrive to the identity
\begin{equation}\label{identity}
 \langle b, b^\dagger\rangle_\nu
=\frac{\langle B, B^\dagger\rangle_\nu+i\langle B b^\dagger\rangle}{(\nu+\mathrm{sgn}(g_{QQ})\nu_0)^2+\kappa^2}-
\frac{i\langle  b b^\dagger\rangle}{\nu+\mathrm{sgn}(g_{QQ})\nu_0-i\kappa}
\end{equation}
Next we replace  the nonlinear spectra in the right hand side of Eq.\  (\ref{identity})
with the spectra of a weakly-damped harmonic oscillator 
\begin{eqnarray}
\langle b^2,b^{\dagger 2}\rangle_\nu^0&\equiv&   \frac{4\kappa(\bar{n}_e+1)^2}{\nu+\mathrm{sgn}(g_{QQ})2\nu_0-i2\kappa}\,,                 \\
\langle b^{\dagger 2}, b^2\rangle_\nu^0&\equiv&   \frac{4\kappa\bar{n}_e^2}{\nu-\mathrm{sgn}(g_{QQ})2\nu_0-i2\kappa}\,,                \\
\langle b^{\dagger }b,b^{\dagger }b\rangle_\nu^0&\equiv& \frac{2\kappa\bar{n}_e(\bar{n}_e+1)}{\nu-i2\kappa}\,,
\end{eqnarray}
and find
\begin{eqnarray}\label{eq:bbd}
\langle b, b^\dagger\rangle_\nu&\approx& \frac{
\kappa (\bar{n}_e+1)}{\nu+\mathrm{sgn}(g_{QQ})\nu_0-i\kappa}+\frac{\lambda V_2^2}{\nu_0^2}\langle b^2,b^{\dagger 2}\rangle_\nu^0\nonumber\\
&&\frac{\lambda V_1^2}{\nu_0^2}\langle b^{\dagger 2}, b^2\rangle_\nu^0
+\frac{\lambda 4 V_2^2}{\nu_0^2}\langle b^{\dagger }b,b^{\dagger }b\rangle_\nu^0\,.
\end{eqnarray}

The result consists of four terms, each of them yielding a Lorentzian peak. 
It is important to keep in mind that the substitution of a nonlinear spectrum with a linear one is valid only close to the peak of the spectrum. Hence, our approximate result for $\langle b, b^\dagger\rangle_\nu$ gives a good approximation close to its second order peaks at  $\pm 2\nu_0,0$, only when the peaks are well pronounced. The peak height, given by Eq.\ (\ref{eq:bbd}), is of the order $\lambda/\kappa$. On the other hand the background value can be computed by linearizing the fluctuations, and is found to be of order
$\kappa$ \cite{Drummond1980c}. This leaves us with the requirement  $\kappa^2\ll\lambda$.

The calculation of the spectra $\langle b^\dagger, b\rangle_\nu$ and  $\langle b, b\rangle_\nu$ 
proceeds along the same lines and yields
\begin{eqnarray}\label{eq:bdb}
\langle b^\dagger, b\rangle_\nu&\approx& \frac{
\kappa \bar{n}_e}{\nu-\mathrm{sgn}(g_{QQ})\nu_0-i\kappa}+\frac{\lambda V_1^2}{\nu_0^2}\langle b^2,b^{\dagger 2}\rangle_\nu^0\nonumber\\
&&+\frac{\lambda V_2^2}{\nu_0^2}\langle b^{\dagger 2}, b^2\rangle_\nu^0
+\frac{\lambda 4 V_2^2}{\nu_0^2}\langle b^{\dagger }b,b^{\dagger }b\rangle_\nu^0\,.
\end{eqnarray}
and
\begin{eqnarray}\label{eq:bb}
  \langle b, b\rangle_\nu&\approx&- \frac{\lambda V_1 V_2}{\nu_0^2}\left(
\langle b^2,b^{\dagger 2}\rangle_\nu^0+\langle b^{\dagger 2}, b^2\rangle_\nu^0\right)\nonumber\\
&&+\frac{\lambda 4 V_2^2}{\nu_0^2}\langle b^{\dagger }b,b^{\dagger }b\rangle_\nu^0\,.
\end{eqnarray}

\bibliographystyle{apsrev4}

\end{document}